\documentclass[preprint,twocolumn,10pt,a4paper,byrevtex,aip,superscriptaddress,showpacs]{revtex4-1}
\usepackage{natbib}
\usepackage{graphicx}
\usepackage{amsfonts}
\usepackage{amsmath}
\usepackage{amssymb}
\usepackage{color}
\usepackage{hyperref}

\begin{document}

\title{Superpoissonian shot noise in organic magnetic tunnel junctions}

\author{Juan Pedro Cascales}
\affiliation{Dpto. Fisica Materia Condensada C3, Instituto Nicolas Cabrera (INC),
Condensed Matter Physics Institute (IFIMAC), Universidad Autonoma de Madrid, Madrid 28049, Spain}

\author{Jhen-Yong Hong}
\affiliation{Department of Physics, National Taiwan University, Taipei 10617, Taiwan and
Institute of Atomic and Molecular Sciences, Academia Sinica, Taipei 10617, Taiwan}

\author{Isidoro Martinez}
\affiliation{Dpto. Fisica Materia Condensada C3, Instituto Nicolas Cabrera (INC),
Condensed Matter Physics Institute (IFIMAC), Universidad Autonoma de Madrid, Madrid 28049, Spain}

\author{Minn-Tsong Lin}
\email{mtlin@phys.ntu.edu.tw}
\affiliation{Department of Physics, National Taiwan University, Taipei 10617, Taiwan and
Institute of Atomic and Molecular Sciences, Academia Sinica, Taipei 10617, Taiwan}

\author{Tomasz Szczepa\'{n}ski}
\affiliation{Department of Physics, Rzesz\'ow University of Technology,
al.~Powsta\'nc\'ow Warszawy 6, 35-959 Rzesz\'ow, Poland}

\author{Vitalii K. Dugaev}
\affiliation{Department of Physics, Rzesz\'ow University of Technology,
al.~Powsta\'nc\'ow Warszawy 6, 35-959 Rzesz\'ow, Poland}

\author{J\'{o}zef Barna\'s}
\affiliation{Faculty of Physics, Adam Mickiewicz University, ul. Umultowska 85, 61-614
Pozna\'n, and Institute of Molecular Physics, Polish Academy of Sciences,
ul. Smoluchowskiego 17, 60-179 Pozna\'n, Poland}

\author{Farkhad G. Aliev*}
\email{(*) Corresponding author: farkhad.aliev@uam.es}
\affiliation{Dpto. Fisica Materia Condensada C3, Instituto Nicolas Cabrera (INC),
Condensed Matter Physics Institute (IFIMAC), Universidad Autonoma de Madrid, Madrid 28049, Spain}

\begin{abstract}
Organic molecules have recently revolutionized ways to create new spintronic devices.
Despite intense studies, the statistics of tunneling electrons through organic barriers remains unclear.
Here we investigate conductance and shot noise in
magnetic tunnel junctions with PTCDA barriers a few nm thick. For junctions in the
electron tunneling regime, with magnetoresistance ratios between 10 and 40\%,
we observe superpoissonian shot noise. The Fano factor exceeds in 1.5-2 times the maximum values reported
for magnetic tunnel junctions with inorganic barriers, indicating spin dependent bunching in tunneling.  We explain
our main findings in terms of a model which includes tunneling through a two level (or multilevel)
system, originated from interfacial bonds of the PTCDA molecules. Our results suggest that
interfaces play an important role in the control of shot noise when electrons tunnel through organic barriers.

\end{abstract}

\maketitle


Organic elements in electronic devices have some advantages over inorganic ones,
like the ability to
chemically adjust their electronic properties, their mechanical flexibility, and the
capability to form self-assembled layers.
Exploring the unique properties of the organic world to improve and create
new functionalities in spin related optics, electronics and memory elements has
been attracting considerable attention in the past decade \cite{Dediu2002,Xiong2004,Rocha2005,Santos2007,Sanvito2011,Jian2011,Vincent2012}.
Organic spintronics may lead to unique devices, for instance, organic light emitting diodes (OLEDs) based on magnetically controlled luminescence \cite{Nguyen2012}.
A key limiting factor for the operation of these and related devices is their signal to noise ratio. Thus, the investigation of noise sources in organic tunnel junctions and spin valves is
of fundamental and technological interest, as the noise ultimately determines their
practical applications.

Low frequency noise, and shot noise (SN) measurements have been systematically used to
characterize the electronic transport mechanisms in inorganic spintronics \cite{Nowak1998,Guerrero2006a,Guerrero2007,Coey2007,Almeida2008,Sekiguchi2010,Arakawa2011,Tanaka2014}.
On the other hand, noise in organic-based devices, which could have $1/f$ and shot
noise contributions, remains poorly understood.
For example, $1/f$ noise measurements have been used to determine
device quality \cite{Clement2007}, or transport features in graphene-based devices
(including one or several layers) \cite{Balandin2013}. In another study,
the $1/f$ noise and DC leakage measurements were used as a diagnostic
tool for OLED reliability in a production line \cite{Rocha2013}.
Current $1/f$ noise measurements have been also used to identify individually contacted
organic molecules \cite{Tsutsui2010,Schaffert2012}.

Earlier noise measurements in organic spintronic devices were carried out at large applied voltages,
where the $1/f$ noise is dominant. Therefore, they were not able to unveil the role of shot
noise, the most fundamental noise source in nanodevices.
Apart from being important from the point of view of applications,
precise knowledge of SN can provide a valuable information on electron correlations
near the interfaces with organic barriers, especially in the regime of direct tunneling.
In fact, the role of interfaces remains one of the central issues in organic spintronics \cite{Keevers2013}.

Here, we analyze the tunneling statistics in organic magnetic tunnel junctions
(O-MTJs) by measuring shot noise, which is known to be an excellent tool to investigate the correlations 
and other details of electron tunneling, well beyond the capabilities of transport measurements 
\cite{BB00,Bulka99,Tserkovnyak2001,Lopez03,Thielmann2003,Cottet04,Belzig2005,Sousa2008,Chudn2008,Yama2011,Schneider2012}.
Being a consequence of the discrete nature of charge carriers, SN is the only contribution to the noise which 
survives down to low temperatures. The normalized shot noise (or Fano factor $F$) indicates \cite{BB00}
whether the tunneling is uncorrelated (poissonian, $F=1$), anti-bunched
(sub-poissonian, typically due to negative correlations, $F<1$) or
bunched (super-poissonian, typically due to positive correlations,
$F>1$).

We have investigated the conductance and shot noise of O-MTJs with
PTCDA molecular barriers in the direct tunneling regime \cite{Li2011}.
In contrast to MTJs with inorganic barriers \cite{Guerrero2006a,Guerrero2007},
tunneling through molecular barriers shows \emph{super-poissonian}
shot noise which additionally depends on the relative alignment of the
electrodes' magnetization. Our observations are accounted for qualitatively
within a model based on spin dependent electron tunneling through an interacting
two-level (or multi-level) system.


The layer sequence in the PTCDA organic spin valves studied in this letter
is: NiFe(25nm)/CoFe(15nm)/Al$_{x}$O(0.6nm)/PTCDA(1.2-5nm)/Al$_{x}$%
O(0.6nm)/CoFe(30nm). The structure was deposited onto a glass substrate, and
the junctions were prepared in a high-vacuum environment with a base
pressure lower than $10^{-8}$ mbar. The metallic layers were deposited by
sputtering with an Ar working pressure of $5\times 10^{-3}$mbar. The PTCDA
layers were grown by thermal evaporation at $10^{-8}$ mbar, with a
deposition rate of 0.1 nm/s. Thin AlO$_{x}$ layers were grown between the
PTCDA layer and both ferromagnetic layers by partially oxidizing Al
in oxygen plasma for 5 s. Figure 1(a) shows a sketch of the
investigated O-MTJs.

The voltage noise was measured using a cross-correlation technique, described
elsewhere\cite{Guerrero2006a,Guerrero2007}. The correct calibration
of our setup has been confirmed by independent studies~\cite{Sekiguchi2010,Arakawa2011}.
The current noise power in the absence of
correlations is Poissonian (full shot noise) and is given by $S_I=2eI$,
where $I$ is the average current and $e$ the electron charge. The voltage full SN is then $S_{full}=2eIR_{d}^{2}$, with $R_{d}$
being the dynamic resistance obtained from the corresponding $I-V$ curves.
We have obtained the experimental SN, $S_{exp}$, by fitting a
Gaussian peak to the histogram of the part of the
spectra independent of frequency (see the Supplemental Material \cite{suppmat}).
The Fano factor $F$ is then calculated as $F=S_{exp}/S_{full}$.

\begin{figure}[tbp]
\begin{center}
\includegraphics[width=8.5cm]
{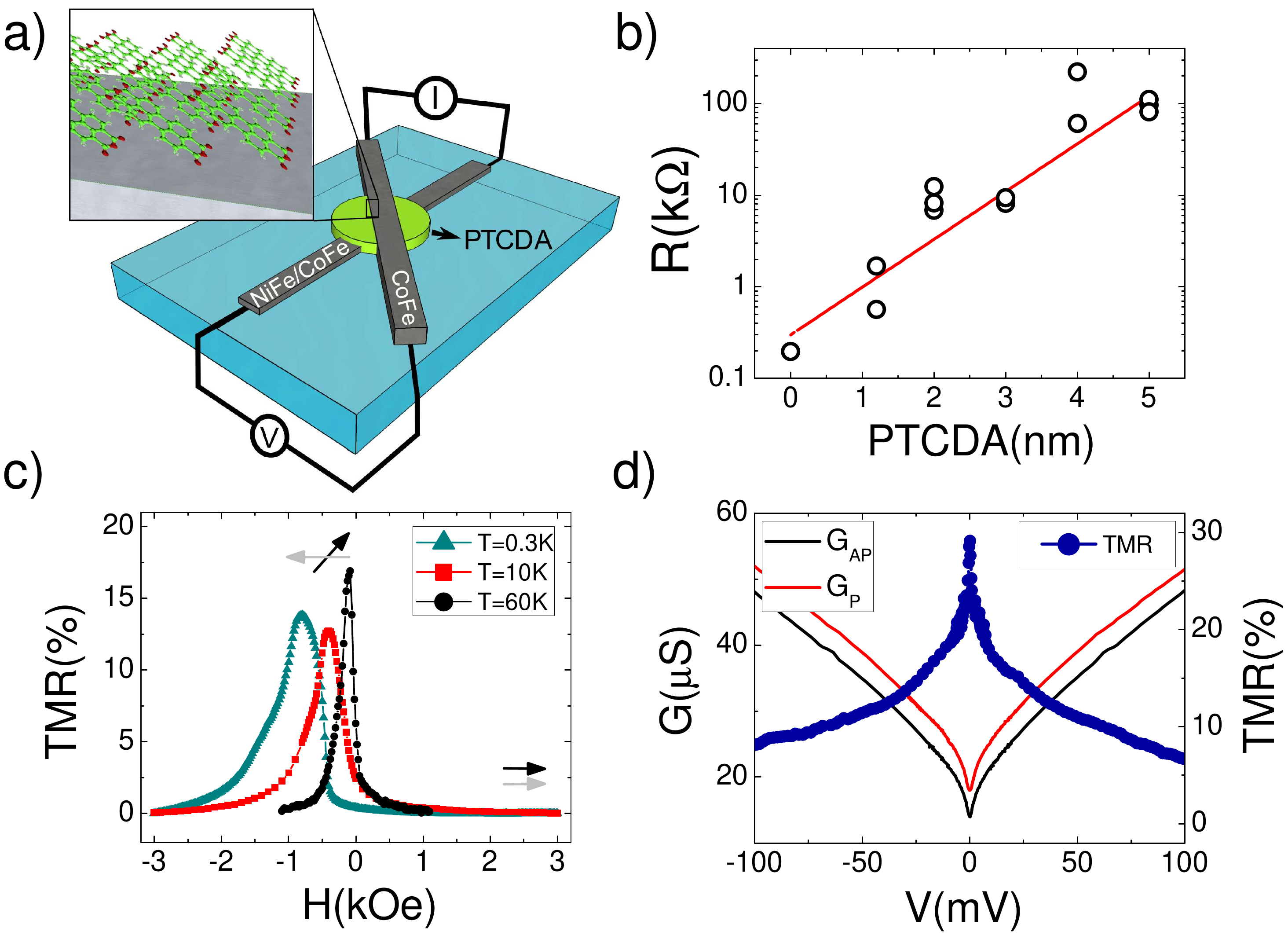}
\end{center}
\caption{(a) Sketch of the sample structure. (b) Experimental dependence of the resistance on the PTCDA
thickness. (c) TMR curves at different temperatures for a sample with 2nm of PTCDA. (d) Dependence of the
TMR and differential conductance on the bias voltage in the P and AP states
for a 4nm PTCDA O-MTJ.}\label{fig:Fig1}
\end{figure}

\begin{figure}[tbp]
\begin{center}
\includegraphics[width=8.5cm]
{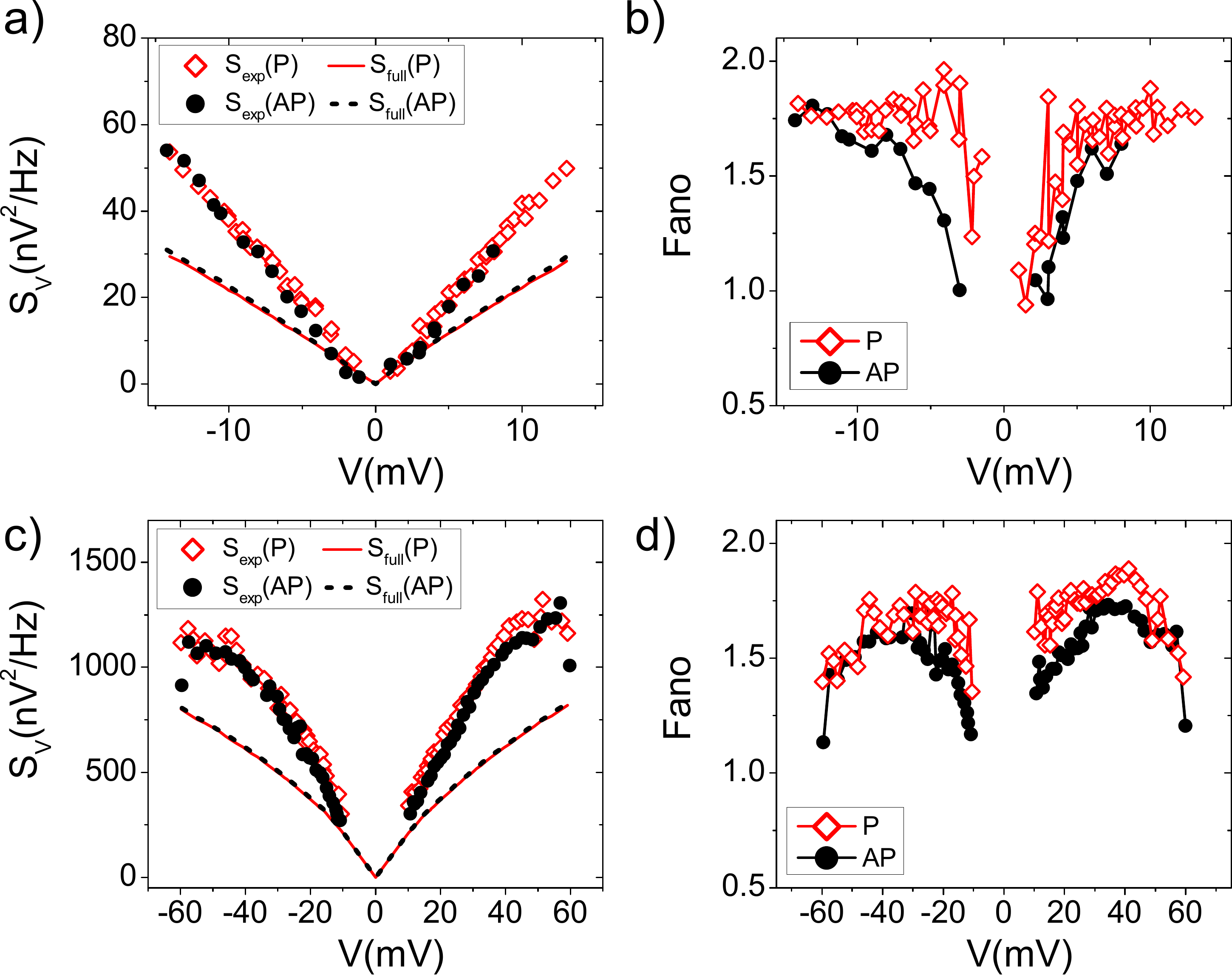}
\end{center}
\caption{Voltage dependence of the noise at T=0.3 K in the P and AP states of the
(a) experimental (dots) and expected full shot noise (lines). (b) Fano factor 
for a 2 nm PTCDA O-MJT. (c) and (d) present similar results for a 5 nm PTCDA O-MJT.}
\end{figure}\label{fig:Fig2}

Figure 1(b) shows that the resistance of the junctions
increases exponentially with the thickness of the PTCDA barrier. This
indicates that the insulating PTCDA layer acts as a barrier in single step  tunneling regime \cite{Schoonus2009}. Figure 1(c) shows the tunneling magnetoresistance (TMR)
for three different temperatures in a 2nm PTCDA O-MTJ, where
the parallel (P) and antiparallel (AP) magnetic alignment of the electrodes are indicated by arrows.
The TMR decreases when the
bias reaches 100 mV (see Fig. 1(d)). Figure 1(d) also presents the differential
conductance in the P and AP states as a function of the bias voltage at
T=0.3K for a 4nm PTCDA O-MTJ. We found that
the magnetic tunnel junctions with PTCDA barriers were more
robust than conventional inorganic MTJs, and typically did not experience dielectrical breakdown
as readily. Out of 14 samples studied,
only 3 have degraded during multiple bias sweeps up to 500mV.

The experimental SN and $S_{full}$ at $T=0.3K$ for
the  2 nm PTCDA junction from Fig. 1(c) are shown
in Fig. 2(a) for the P and AP
states. Fig. 2(b) shows the dependence of the Fano factor with the bias for another
samples with a 5 nm thick PTCDA barrier.
As can be seen, the $F$ factor ranges from $F=1$ at low voltages to $F\simeq 2$ at higher voltages.
All the O-MTJ samples measured displayed a qualitatively similar variation of the Fano factor with the bias voltage.
The shot noise could  be obtained for voltages up to a few tens
of mV only. The maximum voltage for which the shot noise is measured corresponds to
the energy at which the $1/f$ noise becomes dominant and obscures the frequency independent
part of the noise spectrum. Even though the spectra could be obtained up to 100 kHz,  filtering
due to the capacitance of the samples (dependent on the PTCDA thickness)
allowed shot noise measurements only between 1-10 kHz.
The appearance of $1/f$ noise restricted SN measurements
in all the studied samples, especially in the AP state.

Figure 3(a) presents the average
saturation value of the Fano factor in the P state for the samples which
presented the frequency-independent spectra.
Figure 3(a) also shows the variation of TMR with the PTCDA thickness.
Control junctions, with only a 1.2nm AlO$_x$
layer, show TMR below 1\%, and a metallic-like electron transport
(see \cite{suppmat}). This
points to diffusive electron transport,
for which the theory\cite{BB00} predicts the Fano factor equal to 1/3.
Thus, control measurements prove that the super-poissonian SN
is due to the PTCDA barriers. Our O-MTJs with PTCDA thicknesses between 1.2 and
5 nm show relatively high TMR and super-poissonian tunneling statistics with the Fano factor approaching
2, indicative of co-tunneling or tunneling with bunching.
Eight O-MTJs of different barrier thicknesses, from four sample sets,
have shown qualitatively similar SN values (Fig. 3(a)).

A number of electron tunneling mechanisms (Kondo effect\cite{Yamauchi2011},
co-tunneling~\cite{Onac2006,Okazaki2013}, and others \cite{Aguado2007,Kieich2011})
are capable of producing super-poissonian SN. However, they are mostly
relevant for small quantum dots. The observed SN
has been accounted for in terms of the approach developed by Belzig
\cite{Belzig2005}, extended to spin dependent transport.
The corresponding model is based on tunneling through
a two-level system (or multi-level system in a more general case),
with remarkably different tunneling rates through the two levels.
Moreover, these tunneling rates are also spin dependent.
The statistics of the transport process is described as a sum of
independent Poissonian processes transferring bunches of electrons of different size. This arises
from the difference in tunnel rates between the two levels and leads to an enhanced noise.
Details of the model and description will be presented elsewhere.
In Fig. 3(b) we show the TMR and Fano factor in the P and AP states  as a function of the parameter $\beta$, which
describes the spin asymmetry in tunneling rates. The solid and dashed lines are the theoretical results, while the points correspond
to the experimental data. Note that for each sample the TMR and Fano factors have been fitted with the same parameter $\beta$,
which justifies the validity of our approach.

\begin{figure}[tbp]
\hspace*{-0.5cm} \includegraphics[width=8.5cm]{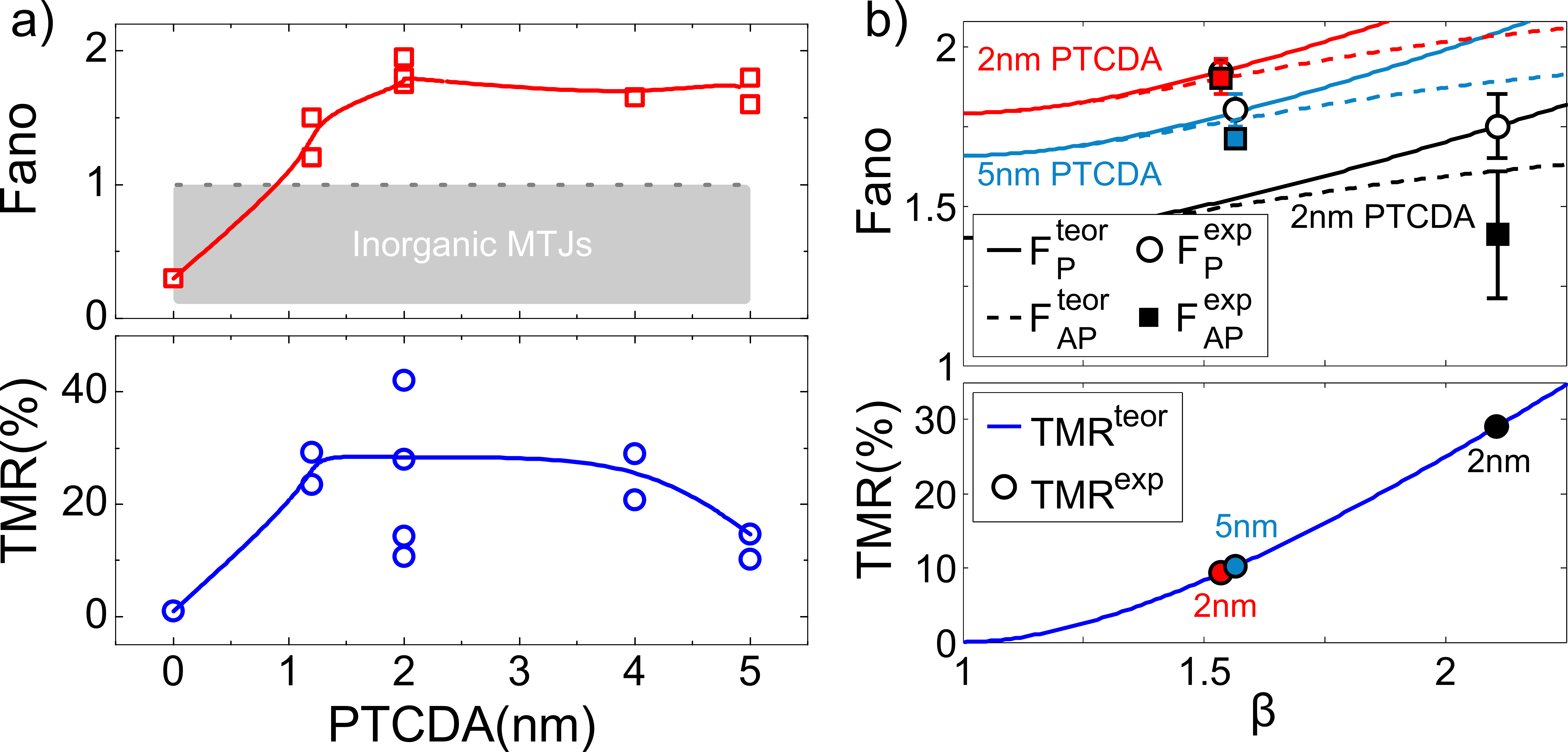}
\caption{(a) Maximum Fano factor in the P state
and the zero bias TMR {\it vs} the PTCDA thickness. The shadowed region corresponds to the range of Fano values for MTJs with inorganic barriers.
(b) Fit of the theory to experimental values of $F$  and  TMR
for the samples  with 2 and 5 nm of PTCDA. The Fano points are obtained from the average saturation value in the Fano factor {\it vs} bias plots, and
the error bars indicate the dispersion of the average.
}\label{fig:fig4}
\end{figure}

Physically, the two or more levels with different couplings, which are responsible for the
observed superpoissonian
shot noise, can have their origin in localized states
arising from interfacial bonds between the PTCDA molecules and the AlO$_x$ buffer layers.
The following arguments suggest that the localized states in the model have an interfacial nature:
(i) the exponential dependence of the tunneling resistance on PTCDA thickness (Fig. 1(b))
including the metallic character of the conductance when only the AlO$_x$ buffer layer is present
(see \cite{suppmat}); (ii) a lateral size of the junctions larger than a micron, for which the influence
of Coulomb blockade is minimized. The physical origin of the interfacial states could be a charge neutrality level 
\cite{Vazquez2004}, or gap states \cite{Yogev2013}, which appear due to the alignment
of the energy levels at metal/organic interfaces \cite{Braun2009}. The bias dependence
of the interfacial density of states could explain the suppression of the Fano factor at large voltages (Fig. 2(b)).

In \emph{conclusion}, super-poissonian statistics in tunneling events through the PTCDA
molecular barriers have been unveiled by shot noise measurements.
A superpoissonian
shot noise has been found, which is likely due to localized states
originated from interfacial bonds of the PTCDA molecules. For a technological
application, the shot noise could be reduced or controlled, for instance, by the growth of double-barrier\cite{Cascales2012} O-MTJs.
Challenges for further work include extending the bias range where the shot noise could be investigated
and comparing the role of the organic layers in the superpoissonian SN by
the study of O-MTJs with different organic layers.

The authors acknowledge support by the Spanish MINECO (MAT2012-32743),
UAM-SANTANDER, Comunidad de Madrid (P2013/MIT2850) and NSC 102-2120-M-002-005 (Taiwan) grants. This work is also partly supported by
the National Center of Research and Development in Poland in the frame of the EU project Era.Net.Rus ``SpinBarrier''.


\begin{thebibliography}{45}
\expandafter\ifx\csname natexlab\endcsname\relax\def\natexlab#1{#1}\fi
\expandafter\ifx\csname bibnamefont\endcsname\relax
  \def\bibnamefont#1{#1}\fi
\expandafter\ifx\csname bibfnamefont\endcsname\relax
  \def\bibfnamefont#1{#1}\fi
\expandafter\ifx\csname citenamefont\endcsname\relax
  \def\citenamefont#1{#1}\fi
\expandafter\ifx\csname url\endcsname\relax
  \def\url#1{\texttt{#1}}\fi
\expandafter\ifx\csname urlprefix\endcsname\relax\def\urlprefix{URL }\fi
\providecommand{\bibinfo}[2]{#2}
\providecommand{\eprint}[2][]{\url{#2}}

\bibitem[{\citenamefont{Dediu et~al.}(2002)\citenamefont{Dediu, Murgia,
  Matacotta, Taliani, and Barbanera}}]{Dediu2002}
\bibinfo{author}{\bibfnamefont{V.}~\bibnamefont{Dediu}},
  \bibinfo{author}{\bibfnamefont{M.}~\bibnamefont{Murgia}},
  \bibinfo{author}{\bibfnamefont{F.}~\bibnamefont{Matacotta}},
  \bibinfo{author}{\bibfnamefont{C.}~\bibnamefont{Taliani}}, \bibnamefont{and}
  \bibinfo{author}{\bibfnamefont{S.}~\bibnamefont{Barbanera}},
  \bibinfo{journal}{Solid State Communications} \textbf{\bibinfo{volume}{122}},
  \bibinfo{pages}{181 } (\bibinfo{year}{2002}), ISSN \bibinfo{issn}{0038-1098}.

\bibitem[{\citenamefont{Xiong et~al.}(2004)\citenamefont{Xiong, Wu,
  Valy~Vardeny, and Shi}}]{Xiong2004}
\bibinfo{author}{\bibfnamefont{Z.~H.} \bibnamefont{Xiong}},
  \bibinfo{author}{\bibfnamefont{D.}~\bibnamefont{Wu}},
  \bibinfo{author}{\bibfnamefont{Z.}~\bibnamefont{Valy~Vardeny}},
  \bibnamefont{and} \bibinfo{author}{\bibfnamefont{J.}~\bibnamefont{Shi}},
  \bibinfo{journal}{Nature} \textbf{\bibinfo{volume}{427}}
  (\bibinfo{year}{2004}).

\bibitem[{\citenamefont{Rocha et~al.}(2005)\citenamefont{Rocha, Garcia-Suarez,
  Bailey, Lambert, Ferrer, and Sanvito}}]{Rocha2005}
\bibinfo{author}{\bibfnamefont{A.~R.} \bibnamefont{Rocha}},
  \bibinfo{author}{\bibfnamefont{V.}~\bibnamefont{Garcia-Suarez}},
  \bibinfo{author}{\bibfnamefont{S.~W.} \bibnamefont{Bailey}},
  \bibinfo{author}{\bibfnamefont{C.~J.} \bibnamefont{Lambert}},
  \bibinfo{author}{\bibfnamefont{J.}~\bibnamefont{Ferrer}}, \bibnamefont{and}
  \bibinfo{author}{\bibfnamefont{S.}~\bibnamefont{Sanvito}},
  \bibinfo{journal}{Nature Materials} \textbf{\bibinfo{volume}{4}},
  \bibinfo{pages}{335} (\bibinfo{year}{2005}).

\bibitem[{\citenamefont{Santos et~al.}(2007)\citenamefont{Santos, Lee, Migdal,
  Lekshmi, Satpati, and Moodera}}]{Santos2007}
\bibinfo{author}{\bibfnamefont{T.~S.} \bibnamefont{Santos}},
  \bibinfo{author}{\bibfnamefont{J.~S.} \bibnamefont{Lee}},
  \bibinfo{author}{\bibfnamefont{P.}~\bibnamefont{Migdal}},
  \bibinfo{author}{\bibfnamefont{I.~C.} \bibnamefont{Lekshmi}},
  \bibinfo{author}{\bibfnamefont{B.}~\bibnamefont{Satpati}}, \bibnamefont{and}
  \bibinfo{author}{\bibfnamefont{J.~S.} \bibnamefont{Moodera}},
  \bibinfo{journal}{Phys. Rev. Lett.} \textbf{\bibinfo{volume}{98}},
  \bibinfo{pages}{016601} (\bibinfo{year}{2007}).

\bibitem[{\citenamefont{Sanvito}(2011)}]{Sanvito2011}
\bibinfo{author}{\bibfnamefont{S.}~\bibnamefont{Sanvito}},
  \bibinfo{journal}{Chemical Society Reviews} \textbf{\bibinfo{volume}{40}}
  (\bibinfo{year}{2011}).

\bibitem[{\citenamefont{Jiang et~al.}(2011)\citenamefont{Jiang, Pearson, and
  Bader}}]{Jian2011}
\bibinfo{author}{\bibfnamefont{J.~S.} \bibnamefont{Jiang}},
  \bibinfo{author}{\bibfnamefont{J.~E.} \bibnamefont{Pearson}},
  \bibnamefont{and} \bibinfo{author}{\bibfnamefont{S.~D.} \bibnamefont{Bader}},
  \bibinfo{journal}{Phys. Rev. Lett.} \textbf{\bibinfo{volume}{106}},
  \bibinfo{pages}{156807} (\bibinfo{year}{2011}).

\bibitem[{\citenamefont{Vincent et~al.}(2012)\citenamefont{Vincent, Klyatskaya,
  Ruben, Wernsdorfer, and Balestro}}]{Vincent2012}
\bibinfo{author}{\bibfnamefont{R.}~\bibnamefont{Vincent}},
  \bibinfo{author}{\bibfnamefont{S.}~\bibnamefont{Klyatskaya}},
  \bibinfo{author}{\bibfnamefont{M.}~\bibnamefont{Ruben}},
  \bibinfo{author}{\bibfnamefont{W.}~\bibnamefont{Wernsdorfer}},
  \bibnamefont{and} \bibinfo{author}{\bibfnamefont{F.}~\bibnamefont{Balestro}},
  \bibinfo{journal}{Nature} \textbf{\bibinfo{volume}{488}}
  (\bibinfo{year}{2012}).

\bibitem[{\citenamefont{Nguyen et~al.}(2012)\citenamefont{Nguyen, Ehrenfreund,
  and Vardeny}}]{Nguyen2012}
\bibinfo{author}{\bibfnamefont{T.~D.} \bibnamefont{Nguyen}},
  \bibinfo{author}{\bibfnamefont{E.}~\bibnamefont{Ehrenfreund}},
  \bibnamefont{and} \bibinfo{author}{\bibfnamefont{Z.~V.}
  \bibnamefont{Vardeny}}, \bibinfo{journal}{Science}
  \textbf{\bibinfo{volume}{337}}, \bibinfo{pages}{204} (\bibinfo{year}{2012}).

\bibitem[{\citenamefont{Nowak et~al.}(1998)\citenamefont{Nowak, Merithew,
  Weissman, Bloom, and Parkin}}]{Nowak1998}
\bibinfo{author}{\bibfnamefont{E.~R.} \bibnamefont{Nowak}},
  \bibinfo{author}{\bibfnamefont{R.~D.} \bibnamefont{Merithew}},
  \bibinfo{author}{\bibfnamefont{M.~B.} \bibnamefont{Weissman}},
  \bibinfo{author}{\bibfnamefont{I.}~\bibnamefont{Bloom}}, \bibnamefont{and}
  \bibinfo{author}{\bibfnamefont{S.~S.~P.} \bibnamefont{Parkin}},
  \bibinfo{journal}{Journal of Applied Physics} \textbf{\bibinfo{volume}{84}}
  (\bibinfo{year}{1998}).

\bibitem[{\citenamefont{Guerrero et~al.}(2006)\citenamefont{Guerrero, Aliev,
  Tserkovnyak, Santos, and Moodera}}]{Guerrero2006a}
\bibinfo{author}{\bibfnamefont{R.}~\bibnamefont{Guerrero}},
  \bibinfo{author}{\bibfnamefont{F.~G.} \bibnamefont{Aliev}},
  \bibinfo{author}{\bibfnamefont{Y.}~\bibnamefont{Tserkovnyak}},
  \bibinfo{author}{\bibfnamefont{T.~S.} \bibnamefont{Santos}},
  \bibnamefont{and} \bibinfo{author}{\bibfnamefont{J.~S.}
  \bibnamefont{Moodera}}, \bibinfo{journal}{Phys. Rev. Lett.}
  \textbf{\bibinfo{volume}{97}}, \bibinfo{pages}{266602}
  (\bibinfo{year}{2006}).

\bibitem[{\citenamefont{Guerrero et~al.}(2007)\citenamefont{Guerrero, Herranz,
  Aliev, Greullet, Tiusan, Hehn, and Montaigne}}]{Guerrero2007}
\bibinfo{author}{\bibfnamefont{R.}~\bibnamefont{Guerrero}},
  \bibinfo{author}{\bibfnamefont{D.}~\bibnamefont{Herranz}},
  \bibinfo{author}{\bibfnamefont{F.~G.} \bibnamefont{Aliev}},
  \bibinfo{author}{\bibfnamefont{F.}~\bibnamefont{Greullet}},
  \bibinfo{author}{\bibfnamefont{C.}~\bibnamefont{Tiusan}},
  \bibinfo{author}{\bibfnamefont{M.}~\bibnamefont{Hehn}}, \bibnamefont{and}
  \bibinfo{author}{\bibfnamefont{F.}~\bibnamefont{Montaigne}},
  \bibinfo{journal}{Applied Physics Letters} \textbf{\bibinfo{volume}{91}},
  \bibinfo{eid}{132504} (\bibinfo{year}{2007}).

\bibitem[{\citenamefont{Scola et~al.}(2007)\citenamefont{Scola, Polovy, Fermon,
  Pannetier-Lecoeur, Feng, Fahy, and Coey}}]{Coey2007}
\bibinfo{author}{\bibfnamefont{J.}~\bibnamefont{Scola}},
  \bibinfo{author}{\bibfnamefont{H.}~\bibnamefont{Polovy}},
  \bibinfo{author}{\bibfnamefont{C.}~\bibnamefont{Fermon}},
  \bibinfo{author}{\bibfnamefont{M.}~\bibnamefont{Pannetier-Lecoeur}},
  \bibinfo{author}{\bibfnamefont{G.}~\bibnamefont{Feng}},
  \bibinfo{author}{\bibfnamefont{K.}~\bibnamefont{Fahy}}, \bibnamefont{and}
  \bibinfo{author}{\bibfnamefont{J.~M.~D.} \bibnamefont{Coey}},
  \bibinfo{journal}{Applied Physics Letters} \textbf{\bibinfo{volume}{90}},
  \bibinfo{eid}{252501} (\bibinfo{year}{2007}).

\bibitem[{\citenamefont{Almeida et~al.}(2008)\citenamefont{Almeida, Wisniowski,
  and Freitas}}]{Almeida2008}
\bibinfo{author}{\bibfnamefont{J.~M.} \bibnamefont{Almeida}},
  \bibinfo{author}{\bibfnamefont{P.}~\bibnamefont{Wisniowski}},
  \bibnamefont{and} \bibinfo{author}{\bibfnamefont{P.}~\bibnamefont{Freitas}},
  \bibinfo{journal}{Magnetics, IEEE Transactions on}
  \textbf{\bibinfo{volume}{44}}, \bibinfo{pages}{2569} (\bibinfo{year}{2008}),
  ISSN \bibinfo{issn}{0018-9464}.

\bibitem[{\citenamefont{Sekiguchi et~al.}(2010)\citenamefont{Sekiguchi,
  Arakawa, Yamauchi, Chida, Yamada, Takahashi, Chiba, Kobayashi, and
  Ono}}]{Sekiguchi2010}
\bibinfo{author}{\bibfnamefont{K.}~\bibnamefont{Sekiguchi}},
  \bibinfo{author}{\bibfnamefont{T.}~\bibnamefont{Arakawa}},
  \bibinfo{author}{\bibfnamefont{Y.}~\bibnamefont{Yamauchi}},
  \bibinfo{author}{\bibfnamefont{K.}~\bibnamefont{Chida}},
  \bibinfo{author}{\bibfnamefont{M.}~\bibnamefont{Yamada}},
  \bibinfo{author}{\bibfnamefont{H.}~\bibnamefont{Takahashi}},
  \bibinfo{author}{\bibfnamefont{D.}~\bibnamefont{Chiba}},
  \bibinfo{author}{\bibfnamefont{K.}~\bibnamefont{Kobayashi}},
  \bibnamefont{and} \bibinfo{author}{\bibfnamefont{T.}~\bibnamefont{Ono}},
  \bibinfo{journal}{Applied Physics Letters} \textbf{\bibinfo{volume}{96}},
  \bibinfo{eid}{252504} (\bibinfo{year}{2010}).

\bibitem[{\citenamefont{Arakawa et~al.}(2011)\citenamefont{Arakawa, Sekiguchi,
  Nakamura, Chida, Nishihara, Chiba, Kobayashi, Fukushima, Yuasa, and
  Ono}}]{Arakawa2011}
\bibinfo{author}{\bibfnamefont{T.}~\bibnamefont{Arakawa}},
  \bibinfo{author}{\bibfnamefont{K.}~\bibnamefont{Sekiguchi}},
  \bibinfo{author}{\bibfnamefont{S.}~\bibnamefont{Nakamura}},
  \bibinfo{author}{\bibfnamefont{K.}~\bibnamefont{Chida}},
  \bibinfo{author}{\bibfnamefont{Y.}~\bibnamefont{Nishihara}},
  \bibinfo{author}{\bibfnamefont{D.}~\bibnamefont{Chiba}},
  \bibinfo{author}{\bibfnamefont{K.}~\bibnamefont{Kobayashi}},
  \bibinfo{author}{\bibfnamefont{A.}~\bibnamefont{Fukushima}},
  \bibinfo{author}{\bibfnamefont{S.}~\bibnamefont{Yuasa}}, \bibnamefont{and}
  \bibinfo{author}{\bibfnamefont{T.}~\bibnamefont{Ono}},
  \bibinfo{journal}{Applied Physics Letters} \textbf{\bibinfo{volume}{98}},
  \bibinfo{eid}{202103} (\bibinfo{year}{2011}).

\bibitem[{\citenamefont{Tanaka et~al.}(2014)\citenamefont{Tanaka, Arakawa,
  Maeda, Kobayashi, Nishihara, Ono, Nozaki, Fukushima, and Yuasa}}]{Tanaka2014}
\bibinfo{author}{\bibfnamefont{T.}~\bibnamefont{Tanaka}},
  \bibinfo{author}{\bibfnamefont{T.}~\bibnamefont{Arakawa}},
  \bibinfo{author}{\bibfnamefont{M.}~\bibnamefont{Maeda}},
  \bibinfo{author}{\bibfnamefont{K.}~\bibnamefont{Kobayashi}},
  \bibinfo{author}{\bibfnamefont{Y.}~\bibnamefont{Nishihara}},
  \bibinfo{author}{\bibfnamefont{T.}~\bibnamefont{Ono}},
  \bibinfo{author}{\bibfnamefont{T.}~\bibnamefont{Nozaki}},
  \bibinfo{author}{\bibfnamefont{A.}~\bibnamefont{Fukushima}},
  \bibnamefont{and} \bibinfo{author}{\bibfnamefont{S.}~\bibnamefont{Yuasa}},
  \bibinfo{journal}{Applied Physics Letters} \textbf{\bibinfo{volume}{105}},
  \bibinfo{eid}{042405} (\bibinfo{year}{2014}).

\bibitem[{\citenamefont{Cl\'ement et~al.}(2007)\citenamefont{Cl\'ement,
  Pleutin, Seitz, Lenfant, and Vuillaume}}]{Clement2007}
\bibinfo{author}{\bibfnamefont{N.}~\bibnamefont{Cl\'ement}},
  \bibinfo{author}{\bibfnamefont{S.}~\bibnamefont{Pleutin}},
  \bibinfo{author}{\bibfnamefont{O.}~\bibnamefont{Seitz}},
  \bibinfo{author}{\bibfnamefont{S.}~\bibnamefont{Lenfant}}, \bibnamefont{and}
  \bibinfo{author}{\bibfnamefont{D.}~\bibnamefont{Vuillaume}},
  \bibinfo{journal}{Phys. Rev. B} \textbf{\bibinfo{volume}{76}},
  \bibinfo{pages}{205407} (\bibinfo{year}{2007}).

\bibitem[{\citenamefont{Balandin}(2013)}]{Balandin2013}
\bibinfo{author}{\bibfnamefont{A.~A.} \bibnamefont{Balandin}},
  \bibinfo{journal}{Nature Nanotechnology} \textbf{\bibinfo{volume}{8}}
  (\bibinfo{year}{2013}).

\bibitem[{\citenamefont{Rocha et~al.}(2013)\citenamefont{Rocha, Gomes,
  Vandamme, De~Leeuw, Meskers, and van~de Weijer}}]{Rocha2013}
\bibinfo{author}{\bibfnamefont{P.}~\bibnamefont{Rocha}},
  \bibinfo{author}{\bibfnamefont{H.}~\bibnamefont{Gomes}},
  \bibinfo{author}{\bibfnamefont{L.}~\bibnamefont{Vandamme}},
  \bibinfo{author}{\bibfnamefont{D.}~\bibnamefont{De~Leeuw}},
  \bibinfo{author}{\bibfnamefont{S.}~\bibnamefont{Meskers}}, \bibnamefont{and}
  \bibinfo{author}{\bibfnamefont{P.}~\bibnamefont{van~de Weijer}}, in
  \emph{\bibinfo{booktitle}{Noise and Fluctuations (ICNF), 2013 22nd
  International Conference on}} (\bibinfo{year}{2013}), pp.
  \bibinfo{pages}{1--4}.

\bibitem[{\citenamefont{Tsutsui}(2010)}]{Tsutsui2010}
\bibinfo{author}{\bibfnamefont{M.~K.~T.} \bibnamefont{Tsutsui},
  \bibfnamefont{Makusu;~Taniguchi}}, \bibinfo{journal}{Nat Commun}
  \textbf{\bibinfo{volume}{1}} (\bibinfo{year}{2010}).

\bibitem[{\citenamefont{Schaffert et~al.}(2012)\citenamefont{Schaffert, Cottin,
  Sonntag, Karacuban, Bobisch, Lorente, Gauyacq, and Muller}}]{Schaffert2012}
\bibinfo{author}{\bibfnamefont{J.}~\bibnamefont{Schaffert}},
  \bibinfo{author}{\bibfnamefont{M.}~\bibnamefont{Cottin}},
  \bibinfo{author}{\bibfnamefont{A.}~\bibnamefont{Sonntag}},
  \bibinfo{author}{\bibfnamefont{H.}~\bibnamefont{Karacuban}},
  \bibinfo{author}{\bibfnamefont{C.}~\bibnamefont{Bobisch}},
  \bibinfo{author}{\bibfnamefont{N.}~\bibnamefont{Lorente}},
  \bibinfo{author}{\bibfnamefont{J.-P.} \bibnamefont{Gauyacq}},
  \bibnamefont{and} \bibinfo{author}{\bibfnamefont{R.}~\bibnamefont{Muller}},
  \bibinfo{journal}{Nature Materials} \textbf{\bibinfo{volume}{12}}
  (\bibinfo{year}{2012}).

\bibitem[{\citenamefont{Keevers et~al.}(2013)\citenamefont{Keevers, Danos,
  Schmidt, and McCamey}}]{Keevers2013}
\bibinfo{author}{\bibfnamefont{T.}~\bibnamefont{Keevers}},
  \bibinfo{author}{\bibfnamefont{A.}~\bibnamefont{Danos}},
  \bibinfo{author}{\bibfnamefont{T.}~\bibnamefont{Schmidt}}, \bibnamefont{and}
  \bibinfo{author}{\bibfnamefont{D.}~\bibnamefont{McCamey}},
  \bibinfo{journal}{Nat. Nanotech.} \textbf{\bibinfo{volume}{8}},
  \bibinfo{pages}{886} (\bibinfo{year}{2013}).

\bibitem[{\citenamefont{Blanter and B�ttiker}(2000)}]{BB00}
\bibinfo{author}{\bibfnamefont{Y.}~\bibnamefont{Blanter}} \bibnamefont{and}
  \bibinfo{author}{\bibfnamefont{M.}~\bibnamefont{B�ttiker}},
  \bibinfo{journal}{Physics Reports} \textbf{\bibinfo{volume}{336}}
  (\bibinfo{year}{2000}).

\bibitem[{\citenamefont{Bu\l{}ka et~al.}(1999)\citenamefont{Bu\l{}ka, Martinek,
  Micha\l{}ek, and Barna\ifmmode~\acute{s}\else \'{s}\fi{}}}]{Bulka99}
\bibinfo{author}{\bibfnamefont{B.~R.} \bibnamefont{Bu\l{}ka}},
  \bibinfo{author}{\bibfnamefont{J.}~\bibnamefont{Martinek}},
  \bibinfo{author}{\bibfnamefont{G.}~\bibnamefont{Micha\l{}ek}},
  \bibnamefont{and}
  \bibinfo{author}{\bibfnamefont{J.}~\bibnamefont{Barna\ifmmode~\acute{s}\else
  \'{s}\fi{}}}, \bibinfo{journal}{Phys. Rev. B} \textbf{\bibinfo{volume}{60}},
  \bibinfo{pages}{12246} (\bibinfo{year}{1999}).

\bibitem[{\citenamefont{Tserkovnyak and Brataas}(2001)}]{Tserkovnyak2001}
\bibinfo{author}{\bibfnamefont{Y.}~\bibnamefont{Tserkovnyak}} \bibnamefont{and}
  \bibinfo{author}{\bibfnamefont{A.}~\bibnamefont{Brataas}},
  \bibinfo{journal}{Phys. Rev. B} \textbf{\bibinfo{volume}{64}},
  \bibinfo{pages}{214402} (\bibinfo{year}{2001}).

\bibitem[{\citenamefont{L\'opez and S\'anchez}(2003)}]{Lopez03}
\bibinfo{author}{\bibfnamefont{R.}~\bibnamefont{L\'opez}} \bibnamefont{and}
  \bibinfo{author}{\bibfnamefont{D.}~\bibnamefont{S\'anchez}},
  \bibinfo{journal}{Phys. Rev. Lett.} \textbf{\bibinfo{volume}{90}},
  \bibinfo{pages}{116602} (\bibinfo{year}{2003}).

\bibitem[{\citenamefont{Thielmann et~al.}(2003)\citenamefont{Thielmann,
  Hettler, K\"onig, and Sch\"on}}]{Thielmann2003}
\bibinfo{author}{\bibfnamefont{A.}~\bibnamefont{Thielmann}},
  \bibinfo{author}{\bibfnamefont{M.~H.} \bibnamefont{Hettler}},
  \bibinfo{author}{\bibfnamefont{J.}~\bibnamefont{K\"onig}}, \bibnamefont{and}
  \bibinfo{author}{\bibfnamefont{G.}~\bibnamefont{Sch\"on}},
  \bibinfo{journal}{Phys. Rev. B} \textbf{\bibinfo{volume}{68}},
  \bibinfo{pages}{115105} (\bibinfo{year}{2003}).

\bibitem[{\citenamefont{Cottet et~al.}(2004)\citenamefont{Cottet, Belzig, and
  Bruder}}]{Cottet04}
\bibinfo{author}{\bibfnamefont{A.}~\bibnamefont{Cottet}},
  \bibinfo{author}{\bibfnamefont{W.}~\bibnamefont{Belzig}}, \bibnamefont{and}
  \bibinfo{author}{\bibfnamefont{C.}~\bibnamefont{Bruder}},
  \bibinfo{journal}{Phys. Rev. Lett.} \textbf{\bibinfo{volume}{92}},
  \bibinfo{pages}{206801} (\bibinfo{year}{2004}).

\bibitem[{\citenamefont{Belzig}(2005)}]{Belzig2005}
\bibinfo{author}{\bibfnamefont{W.}~\bibnamefont{Belzig}},
  \bibinfo{journal}{Phys. Rev. B} \textbf{\bibinfo{volume}{71}},
  \bibinfo{pages}{161301} (\bibinfo{year}{2005}).

\bibitem[{\citenamefont{Souza et~al.}(2008)\citenamefont{Souza, Jauho, and
  Egues}}]{Sousa2008}
\bibinfo{author}{\bibfnamefont{F.~M.} \bibnamefont{Souza}},
  \bibinfo{author}{\bibfnamefont{A.~P.} \bibnamefont{Jauho}}, \bibnamefont{and}
  \bibinfo{author}{\bibfnamefont{J.~C.} \bibnamefont{Egues}},
  \bibinfo{journal}{Phys. Rev. B} \textbf{\bibinfo{volume}{78}},
  \bibinfo{pages}{155303} (\bibinfo{year}{2008}).

\bibitem[{\citenamefont{Chudnovskiy et~al.}(2008)\citenamefont{Chudnovskiy,
  Swiebodzinski, and Kamenev}}]{Chudn2008}
\bibinfo{author}{\bibfnamefont{A.~L.} \bibnamefont{Chudnovskiy}},
  \bibinfo{author}{\bibfnamefont{J.}~\bibnamefont{Swiebodzinski}},
  \bibnamefont{and} \bibinfo{author}{\bibfnamefont{A.}~\bibnamefont{Kamenev}},
  \bibinfo{journal}{Phys. Rev. Lett.} \textbf{\bibinfo{volume}{101}},
  \bibinfo{pages}{066601} (\bibinfo{year}{2008}).

\bibitem[{\citenamefont{Yamauchi
  et~al.}(2011{\natexlab{a}})\citenamefont{Yamauchi, Sekiguchi, Chida, Arakawa,
  Nakamura, Kobayashi, Ono, Fujii, and Sakano}}]{Yama2011}
\bibinfo{author}{\bibfnamefont{Y.}~\bibnamefont{Yamauchi}},
  \bibinfo{author}{\bibfnamefont{K.}~\bibnamefont{Sekiguchi}},
  \bibinfo{author}{\bibfnamefont{K.}~\bibnamefont{Chida}},
  \bibinfo{author}{\bibfnamefont{T.}~\bibnamefont{Arakawa}},
  \bibinfo{author}{\bibfnamefont{S.}~\bibnamefont{Nakamura}},
  \bibinfo{author}{\bibfnamefont{K.}~\bibnamefont{Kobayashi}},
  \bibinfo{author}{\bibfnamefont{T.}~\bibnamefont{Ono}},
  \bibinfo{author}{\bibfnamefont{T.}~\bibnamefont{Fujii}}, \bibnamefont{and}
  \bibinfo{author}{\bibfnamefont{R.}~\bibnamefont{Sakano}},
  \bibinfo{journal}{Phys. Rev. Lett.} \textbf{\bibinfo{volume}{106}},
  \bibinfo{pages}{176601} (\bibinfo{year}{2011}{\natexlab{a}}).

\bibitem[{\citenamefont{Schneider et~al.}(2012)\citenamefont{Schneider, L\"u,
  Brandbyge, and Berndt}}]{Schneider2012}
\bibinfo{author}{\bibfnamefont{N.~L.} \bibnamefont{Schneider}},
  \bibinfo{author}{\bibfnamefont{J.~T.} \bibnamefont{L\"u}},
  \bibinfo{author}{\bibfnamefont{M.}~\bibnamefont{Brandbyge}},
  \bibnamefont{and} \bibinfo{author}{\bibfnamefont{R.}~\bibnamefont{Berndt}},
  \bibinfo{journal}{Phys. Rev. Lett.} \textbf{\bibinfo{volume}{109}},
  \bibinfo{pages}{186601} (\bibinfo{year}{2012}).

\bibitem[{\citenamefont{Li et~al.}(2011)\citenamefont{Li, Chang, Agilan, Hong,
  Tai, Chiang, Fukutani, Dowben, and Lin}}]{Li2011}
\bibinfo{author}{\bibfnamefont{K.-S.} \bibnamefont{Li}},
  \bibinfo{author}{\bibfnamefont{Y.-M.} \bibnamefont{Chang}},
  \bibinfo{author}{\bibfnamefont{S.}~\bibnamefont{Agilan}},
  \bibinfo{author}{\bibfnamefont{J.-Y.} \bibnamefont{Hong}},
  \bibinfo{author}{\bibfnamefont{J.-C.} \bibnamefont{Tai}},
  \bibinfo{author}{\bibfnamefont{W.-C.} \bibnamefont{Chiang}},
  \bibinfo{author}{\bibfnamefont{K.}~\bibnamefont{Fukutani}},
  \bibinfo{author}{\bibfnamefont{P.~A.} \bibnamefont{Dowben}},
  \bibnamefont{and} \bibinfo{author}{\bibfnamefont{M.-T.} \bibnamefont{Lin}},
  \bibinfo{journal}{Phys. Rev. B} \textbf{\bibinfo{volume}{83}},
  \bibinfo{pages}{172404} (\bibinfo{year}{2011}).

\bibitem[{sup()}]{suppmat}
\emph{\bibinfo{title}{See the {S}upplemental {M}aterial for details about the
  estimation of shot noise and the dependence of the conductance with the
  {PTCDA} thickness.}}

\bibitem[{\citenamefont{Schoonus et~al.}(2009)\citenamefont{Schoonus, Lumens,
  Wagemans, Kohlhepp, Bobbert, Swagten, and Koopmans}}]{Schoonus2009}
\bibinfo{author}{\bibfnamefont{J.~J. H.~M.} \bibnamefont{Schoonus}},
  \bibinfo{author}{\bibfnamefont{P.~G.~E.} \bibnamefont{Lumens}},
  \bibinfo{author}{\bibfnamefont{W.}~\bibnamefont{Wagemans}},
  \bibinfo{author}{\bibfnamefont{J.~T.} \bibnamefont{Kohlhepp}},
  \bibinfo{author}{\bibfnamefont{P.~A.} \bibnamefont{Bobbert}},
  \bibinfo{author}{\bibfnamefont{H.~J.~M.} \bibnamefont{Swagten}},
  \bibnamefont{and} \bibinfo{author}{\bibfnamefont{B.}~\bibnamefont{Koopmans}},
  \bibinfo{journal}{Phys. Rev. Lett.} \textbf{\bibinfo{volume}{103}},
  \bibinfo{pages}{146601} (\bibinfo{year}{2009}).

\bibitem[{\citenamefont{Yamauchi
  et~al.}(2011{\natexlab{b}})\citenamefont{Yamauchi, Sekiguchi, Chida, Arakawa,
  Nakamura, Kobayashi, Ono, Fujii, and Sakano}}]{Yamauchi2011}
\bibinfo{author}{\bibfnamefont{Y.}~\bibnamefont{Yamauchi}},
  \bibinfo{author}{\bibfnamefont{K.}~\bibnamefont{Sekiguchi}},
  \bibinfo{author}{\bibfnamefont{K.}~\bibnamefont{Chida}},
  \bibinfo{author}{\bibfnamefont{T.}~\bibnamefont{Arakawa}},
  \bibinfo{author}{\bibfnamefont{S.}~\bibnamefont{Nakamura}},
  \bibinfo{author}{\bibfnamefont{K.}~\bibnamefont{Kobayashi}},
  \bibinfo{author}{\bibfnamefont{T.}~\bibnamefont{Ono}},
  \bibinfo{author}{\bibfnamefont{T.}~\bibnamefont{Fujii}}, \bibnamefont{and}
  \bibinfo{author}{\bibfnamefont{R.}~\bibnamefont{Sakano}},
  \bibinfo{journal}{Phys. Rev. Lett.} \textbf{\bibinfo{volume}{106}},
  \bibinfo{pages}{176601} (\bibinfo{year}{2011}{\natexlab{b}}).

\bibitem[{\citenamefont{Onac et~al.}(2006)\citenamefont{Onac, Balestro,
  Trauzettel, Lodewijk, and Kouwenhoven}}]{Onac2006}
\bibinfo{author}{\bibfnamefont{E.}~\bibnamefont{Onac}},
  \bibinfo{author}{\bibfnamefont{F.}~\bibnamefont{Balestro}},
  \bibinfo{author}{\bibfnamefont{B.}~\bibnamefont{Trauzettel}},
  \bibinfo{author}{\bibfnamefont{C.~F.~J.} \bibnamefont{Lodewijk}},
  \bibnamefont{and} \bibinfo{author}{\bibfnamefont{L.~P.}
  \bibnamefont{Kouwenhoven}}, \bibinfo{journal}{Phys. Rev. Lett.}
  \textbf{\bibinfo{volume}{96}}, \bibinfo{pages}{026803}
  (\bibinfo{year}{2006}).

\bibitem[{\citenamefont{Okazaki et~al.}(2013)\citenamefont{Okazaki, Sasaki, and
  Muraki}}]{Okazaki2013}
\bibinfo{author}{\bibfnamefont{Y.}~\bibnamefont{Okazaki}},
  \bibinfo{author}{\bibfnamefont{S.}~\bibnamefont{Sasaki}}, \bibnamefont{and}
  \bibinfo{author}{\bibfnamefont{K.}~\bibnamefont{Muraki}},
  \bibinfo{journal}{Phys. Rev. B} \textbf{\bibinfo{volume}{87}},
  \bibinfo{pages}{041302} (\bibinfo{year}{2013}).

\bibitem[{\citenamefont{Lambert et~al.}(2007)\citenamefont{Lambert, Aguado, and
  Brandes}}]{Aguado2007}
\bibinfo{author}{\bibfnamefont{N.}~\bibnamefont{Lambert}},
  \bibinfo{author}{\bibfnamefont{R.}~\bibnamefont{Aguado}}, \bibnamefont{and}
  \bibinfo{author}{\bibfnamefont{T.}~\bibnamefont{Brandes}},
  \bibinfo{journal}{Phys. Rev. B} \textbf{\bibinfo{volume}{75}},
  \bibinfo{pages}{045340} (\bibinfo{year}{2007}).

\bibitem[{\citenamefont{Kie\ss{}lich et~al.}(2007)\citenamefont{Kie\ss{}lich,
  Sch\"oll, Brandes, Hohls, and Haug}}]{Kieich2011}
\bibinfo{author}{\bibfnamefont{G.}~\bibnamefont{Kie\ss{}lich}},
  \bibinfo{author}{\bibfnamefont{E.}~\bibnamefont{Sch\"oll}},
  \bibinfo{author}{\bibfnamefont{T.}~\bibnamefont{Brandes}},
  \bibinfo{author}{\bibfnamefont{F.}~\bibnamefont{Hohls}}, \bibnamefont{and}
  \bibinfo{author}{\bibfnamefont{R.~J.} \bibnamefont{Haug}},
  \bibinfo{journal}{Phys. Rev. Lett.} \textbf{\bibinfo{volume}{99}},
  \bibinfo{pages}{206602} (\bibinfo{year}{2007}).

\bibitem[{\citenamefont{V\'azquez et~al.}(2004)\citenamefont{V\'azquez,
  Oszwaldowski, Pou, Ortega, P\'erez, Flores, and Kahn}}]{Vazquez2004}
\bibinfo{author}{\bibfnamefont{H.}~\bibnamefont{V\'azquez}},
  \bibinfo{author}{\bibfnamefont{R.}~\bibnamefont{Oszwaldowski}},
  \bibinfo{author}{\bibfnamefont{P.}~\bibnamefont{Pou}},
  \bibinfo{author}{\bibfnamefont{J.}~\bibnamefont{Ortega}},
  \bibinfo{author}{\bibfnamefont{R.}~\bibnamefont{P\'erez}},
  \bibinfo{author}{\bibfnamefont{F.}~\bibnamefont{Flores}}, \bibnamefont{and}
  \bibinfo{author}{\bibfnamefont{A.}~\bibnamefont{Kahn}}, \bibinfo{journal}{EPL
  (Europhysics Letters)} \textbf{\bibinfo{volume}{65}}, \bibinfo{pages}{802}
  (\bibinfo{year}{2004}).

\bibitem[{\citenamefont{Yogev et~al.}(2013)\citenamefont{Yogev, Matsubara,
  Nakamura, Zschieschang, Klauk, and Rosenwaks}}]{Yogev2013}
\bibinfo{author}{\bibfnamefont{S.}~\bibnamefont{Yogev}},
  \bibinfo{author}{\bibfnamefont{R.}~\bibnamefont{Matsubara}},
  \bibinfo{author}{\bibfnamefont{M.}~\bibnamefont{Nakamura}},
  \bibinfo{author}{\bibfnamefont{U.}~\bibnamefont{Zschieschang}},
  \bibinfo{author}{\bibfnamefont{H.}~\bibnamefont{Klauk}}, \bibnamefont{and}
  \bibinfo{author}{\bibfnamefont{Y.}~\bibnamefont{Rosenwaks}},
  \bibinfo{journal}{Phys. Rev. Lett.} \textbf{\bibinfo{volume}{110}},
  \bibinfo{pages}{036803} (\bibinfo{year}{2013}).

\bibitem[{\citenamefont{Braun et~al.}(2009)\citenamefont{Braun, Salaneck, and
  Fahlman}}]{Braun2009}
\bibinfo{author}{\bibfnamefont{S.}~\bibnamefont{Braun}},
  \bibinfo{author}{\bibfnamefont{W.~R.} \bibnamefont{Salaneck}},
  \bibnamefont{and} \bibinfo{author}{\bibfnamefont{M.}~\bibnamefont{Fahlman}},
  \bibinfo{journal}{Advanced Materials} \textbf{\bibinfo{volume}{21}},
  \bibinfo{pages}{1450} (\bibinfo{year}{2009}), ISSN \bibinfo{issn}{1521-4095}.

\bibitem[{\citenamefont{Cascales et~al.}(2012)\citenamefont{Cascales, Herranz,
  Aliev, Szczepa\ifmmode~\acute{n}\else \'{n}\fi{}ski, Dugaev,
  Barna\ifmmode~\acute{s}\else \'{s}\fi{}, Duluard, Hehn, and
  Tiusan}}]{Cascales2012}
\bibinfo{author}{\bibfnamefont{J.~P.} \bibnamefont{Cascales}},
  \bibinfo{author}{\bibfnamefont{D.}~\bibnamefont{Herranz}},
  \bibinfo{author}{\bibfnamefont{F.~G.} \bibnamefont{Aliev}},
  \bibinfo{author}{\bibfnamefont{T.}~\bibnamefont{Szczepa\ifmmode~\acute{n}\else
  \'{n}\fi{}ski}}, \bibinfo{author}{\bibfnamefont{V.~K.} \bibnamefont{Dugaev}},
  \bibinfo{author}{\bibfnamefont{J.}~\bibnamefont{Barna\ifmmode~\acute{s}\else
  \'{s}\fi{}}}, \bibinfo{author}{\bibfnamefont{A.}~\bibnamefont{Duluard}},
  \bibinfo{author}{\bibfnamefont{M.}~\bibnamefont{Hehn}}, \bibnamefont{and}
  \bibinfo{author}{\bibfnamefont{C.}~\bibnamefont{Tiusan}},
  \bibinfo{journal}{Phys. Rev. Lett.} \textbf{\bibinfo{volume}{109}},
  \bibinfo{pages}{066601} (\bibinfo{year}{2012}).

\end{thebibliography}
\end{document}